**We must re-evaluate assumptions about carbon trading for effective climate change mitigation**

Alyssa R. Pfadt-Trilling[a], Marie-Odile P. Fortier[b]*

a. Environmental Systems Graduate Group, University of California, Merced, 5200 North Lake Road, Merced, CA 95343, USA.

b. Department of Civil and Environmental Engineering and Construction, University of Nevada, Las Vegas, 4505 South Maryland Parkway, Las Vegas, NV 89154, USA.

*Corresponding author, marie-odile.fortier@unlv.edu

**Abstract:**
Climate policy decision-makers are generally provided with simplified information that overlooks the complexities of the climate crisis. The success of current and planned climate change mitigation strategies depends upon the assumption that $CO_2$ and other greenhouse gases (GHGs) are fungible: that the 'value' or potential to induce atmospheric warming of one unit of $CO_2$ is origin- and path-independent. This assumption has been ubiquitously integrated into key policy despite previous challenges in the scientific literature. We address how the uncertainties in normalizing non-$CO_2$ GHG species to $CO_2$-equivalents are obscured by reporting single-point values for GHG accounting. We demonstrate that the foundation of emissions trading, the assumption that different sources and sinks of $CO_2$ emissions are interchangeable in how they contribute to climate change, is scientifically unsound. We discuss numerous failures of carbon offset programs, and how consequently, net GHG emissions have increased. Evidence indicates that treating carbon as a fungible commodity is unlikely to produce the intended outcome of mitigating climate change. We recommend that decarbonization plans not rely on carbon offsetting, that differentiated data be used instead of $CO_2$-equivalents by decision-makers, and that emission reduction targets be separated by GHG species to achieve long-term climate stabilization.

1. **Introduction**

Genuine solutions for reducing anthropogenic greenhouse gas (GHG) emissions are urgently required to have any chance of meeting major climate targets. Climate change mitigation will not be successful if we prioritize and enact actions that are not demonstrably effective, both in their physical science basis and in the way they are implemented and managed. The first Intergovernmental Panel on Climate Change (IPCC) report recommended significant emissions reductions including immediately reducing $CO_2$ by 60% from 1990 levels, that have still not materialized as of 2025.[1,2] Instead, the rate of emissions has accelerated; the atmospheric concentration of $CO_2$ has increased just as much between 1990 and 2023 as it had from the pre-industrial era to 1990 (~70 $ppm_v$).[3,4] While the increasing atmospheric concentration of carbon dioxide ($CO_2$) is the predominant cause of climate change, it is understood that other GHG species and changes to the Earth's surface reflectivity also contribute to the climate crisis.[5] GHGs including methane ($CH_4$), nitrous oxide ($N_2O$), and others (e.g. chlorofluorocarbons or CFCs) all contribute to the greenhouse effect, although different GHGs have varying atmospheric residence times and absorb different bandwidths, resulting in different climatic impacts. The complexity of these processes has necessitated simplifying assumptions to communicate and plan policy. Climate policy has consequently been shaped by these assumptions, despite serious consequences if their use propagates and magnifies their inaccuracies.[6]

The basis of mainstream climate policy assumes that all GHGs are fungible in the form of static single-point $CO_2$ equivalent mass units ($CO_2$eq). Fungibility enables straightforward trading of a commodity; the value of a fungible good is origin- and path-independent. Currencies are fungible by definition; the purchasing power of money depends only on the denomination, and so cash has the same value regardless of where or how those bills have been previously used. In contrast, units of a nonfungible commodity like land assets are specific to a unit. A hectare of land does not necessarily have equal value to another hectare of land in a different location, and two half-hectare plots do not hold the same value as a single-hectare plot.

Decarbonization plans currently rely on trading the climatic impact or value of GHGs through systems of carbon credits and offsets as if all GHGs are perfectly fungible. Carbon offsets include reduced deforestation, reforestation, afforestation, deployment of renewable energy as a substitute for fossil fuel use, industrial refrigerant destruction, soil amendments or altered practices to increase soil organic carbon content, and direct air capture of $CO_2$, among other approaches. Because $CO_2$ is a well-mixed GHG with an extremely large perturbation time,[5] it has been posited that emitting one ton of $CO_2$ anywhere on earth has the same climatic impact as emitting one ton of $CO_2$ anywhere else.[7] However, this has been misconstrued in policy, as the climatic impact of $CO_2$ *absorbed* in different regions and by different mechanisms is not equal. This assumption that facilitates carbon trading and tracking of emissions has also led to the misconception that $CO_2$ emissions from the use of fossil fuels can be negated by carbon offsets, as well as the conflation of reductions in rates of GHG emissions with actual removals or sequestration of GHGs.

Although the climate science literature has long demonstrated these complexities,[8–10] and the social science literature has heavily criticized the idea of commodifying carbon,[11–15] these challenges have not been systematically incorporated into climate action planning in practice. The US government has shown support for carbon trading as a key method for decarbonization in their report "Voluntary Carbon Markets Joint Policy Statement and Principles" released in May 2024, which does not address all of the issues with carbon credits.[16] In this work, we investigate key assumptions used in developing climate change mitigation strategies and explore their consequences to provide guidance for decision-makers.

## 2. Assumption #1: Fungibility of Greenhouse Gases via the Global Warming Potential

The practice of treating the climatic impact of different GHG species as fungible first appeared in public policy in the Kyoto Protocol, with the Global Warming Potential (GWP) equation used to include multiple gases within the treaty.[17] The GWP equation (Equation 1) provides a relative contribution towards climate change in $CO_2$eq for a GHG species of interest ($i$) by integrating the instantaneous radiative forcing ($a$) resulting from one additional unit increase in species $i$ and its concentration ($c$) remaining at time $t$ over a chosen time horizon $n$.[18]

$$GWP = \frac{\int_0^n a_i c_i dt}{\int_0^n a_{CO_2} c_{CO_2} dt} \qquad \text{(Equation 1)}$$

The Kyoto Protocol involved years of negotiations, during which time major concessions were made. When the Kyoto Protocol entered into force in 2005, it was based on information from the 1990 IPCC report, which had already been updated twice by then. This first IPCC report included 20, 100, and 500 years as arbitrary options for the time horizon through which to compare the GWPs of various GHGs as $CO_2$eq.[19] The standard option at the Kyoto Protocol conference became 100 years, simply because it was the middle option presented in the IPCC report, inadvertently setting a policy standard since then.[17] However, there has never been a strong scientific consensus that the climatic impact of all GHGs should be related in terms of $CO_2$eq, and one of the original authors later wrote that the GWP was not intended to be used for policy or as a universally accepted standard.[2,17] In 1997, Skodvin and Fuglestvedt described the GWP as a "preliminary tool for policymaking until better methods are developed."[20]

The GWP equation includes two major limitations that were originally reported: that the effective radiative forcing of a GHG depends on atmospheric composition, including the lifetime and concentration of that gas, and that the lifetime and effects of $CO_2$ are highly uncertain.[2,18] Because the GWP of a GHG relative to $CO_2$ depends on atmospheric concentration, it is a dynamic value and is updated in every IPCC report since 2001 (Figure 1).[2,21,22] Thus, these GWPs values have changed over time with the evolving scientific understanding of different GHG lifetimes and efficacies as well as with the changes to atmospheric concentrations since reporting began. As GWP depends on atmospheric composition, the GWP values also change under different representative concentration pathways (RCPs), thus limiting its utility in climate policy. The projections of different RCPs in turn change the GWPs of GHGs; for example, methane GWPs increase under the lowest pathway and decrease under the highest pathway.[21]

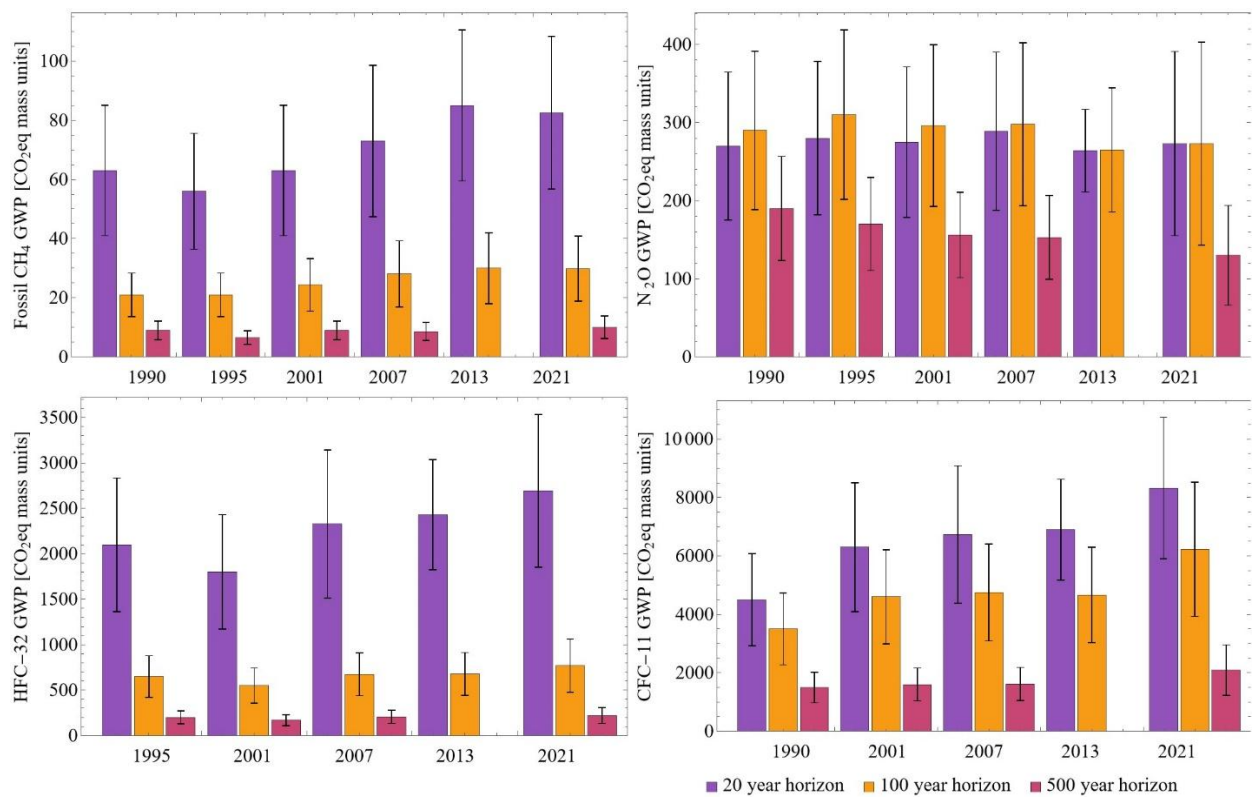

**Figure 1:** The range of 100-year global warming potential (GWP) values for methane (CH$_4$), nitrous oxide (N$_2$O), difluoromethane (HFC-32), and trichlorofluoromethane (CFC-11) relative to CO$_2$ reported in each of the six IPCC reports from 1990 to 2021, including the uncertainties of ±35% reported from 1990 through 2007, ±40% in 2013, and ±11% in the latest IPCC report in 2021. In the fifth report in 2013, the IPCC began differentiating between biogenic and fossil-based methane, with fossil-based methane values also accounting for +2 kg CO$_2$ molecules that stay in the atmosphere after oxidation.[2,5,22–25]

Since the first IPCC report, the consensus has shifted on the estimated lifetime of CO$_2$ in the atmosphere; we now understand that the perturbation time of CO$_2$ is significantly longer than the residence time, and the level of uncertainty makes it inappropriate to assign it a single value.[2,5] Furthermore, aggregating GHGs and reporting their GWP relative to CO$_2$ involves considerable uncertainty that is not accurately represented in current practices. Each IPCC report contains an uncertainty range for the 95% confidence level for relating different GHGs to CO$_2$eq (Figure 1), but that uncertainty is obscured by aggregating climate change impacts together. It has been previously argued that the uncertainty associated with the GWP index renders it irrelevant for policy applications.[26]

There have been challenges to using GWP in climate policy since shortly after the first IPCC report due to appearing more scientifically sound than they really are.[19] One of the main criticisms of using GWP for policy purposes is that aggregated emissions in CO$_2$-equivalents do not actually lead to the same estimated temperature outcomes over time.[5] The GWP equation obscures differences in the impacts of short-lived climate forcers (SLCFs) and long-lived, well-mixed GHGs.[27–30] As SLCFs do not persist in the atmosphere for extended periods of time, their long-term impact on climate stabilization could potentially be misrepresented or misinterpreted when expressed as $CO_2$eq using the GWP equation. The fact that the warming potential of SLCFs like methane depends on the rate of emissions means that, theoretically, temperatures can be stabilized without reaching net-zero methane, as opposed to the cumulative effect of carbon dioxide, which must reach net-zero emissions in order to halt warming.[31,32] Prioritizing reducing SLCFs like methane, versus the equivalent amount of $CO_2$ according to the 100-year GWP, would result in very different climatic outcomes, both in the rate of temperature change and absolute temperature increase.[33] Aggregating SLCFs and $CO_2$ using the 100-year GWP to meet peak warming targets could lead to overshoot, where average temperatures temporarily increase above 2°C compared to pre-industrial average temperature and then hypothetically reduce and stabilize.[32] Overshoot is considered very risky and may result in irreparable damage to ecosystems.[34]

The use of separate emissions metrics and policy targets for long-lived versus short-lived GHG species has been proposed as one solution.[32,35] Tanaka and O'Neill (2018) recommend focusing on net-zero CO2 rather than net-zero GHGs.[36] Alternatively, multiple new metrics have been suggested to improve upon the GWP approach (Table 1). In the calculations used in these alternative metrics, the value of non-$CO_2$ GHGs relative to $CO_2$ varies significantly; for example, $CO_2$eq estimates for methane range from 4 to 199 g $CO_2$eq/g $CH_4$ across metrics.[37] None of these alternatives have been widely adopted or included in policy efforts at the time of this writing, despite continued development in metrics, particularly the GWP*.[38,39]

**Table 1**: Overview of alternative metrics to GWP proposed in the scientific literature (non-exhaustive list).

| Alternative metric | Citation | Purpose | Suggested utilization |
| --- | --- | --- | --- |
| $GTP_p$ & $GTP_s$ | Shine et al. (2005)[40] | Represent Global Temperature Change Potential at a given time from a pulse of GHG emissions and the effect of sustained emissions | Comparing the effects of a pulse and sustained emissions; a general replacement of GWP |
| Time Adjusted Warming Potential (TAWP) | Kendall (2011)[41] | Adjusts the efficacy of GHGs according to the timing of release | Projects that occur over an extended time period, or comparison of alternatives over an extended time period |
| Absolute Peak Commitment Temperature & Absolute Sustained Emission Temperature (aPCT & aSET) | Smith et al. (2012)[32] | Measures the temperature change potential from sustained emissions (as opposed to pulse) | Endpoint metric |
| Global Precipitation-change Potential from pulse or sustained emissions ($GPP_p$ & $GPP_s$) | Shine et al. (2015)[42] | Measures potential changes to precipitation instead of temperature | Provides additional context to be used alongside GWP for greater understanding of emissions impacts |
| Sustained-flux GWP & Sustained-flux global cooling potential (SGWP & SGCP) | Neubauer & Megonigal (2015)[30] | Differentiates between gas emissions and gas uptake | Intended to determine whether different ecosystems have a net cooling or net warming effect |
| GWP* | Allen et al. (2018)[27] | Takes into account the difference in cumulative emissions effects of short and long-lived climate forcers | "Benefits of GWP* are most apparent when SLCP emission rates are declining" |
| GWP* | Cain et al. (2019)[39] | | |

| Combined GWP & Combined Global Temperature Change Potential | Collins et al. (2020)[43] | Endpoint metric | Builds on GWP* to combine step and pulse emissions into a 'single basket' endpoint metric for policy |
|---|---|---|---|
| Modified GWP | Abernethy & Jackson (2022)[44] | Suggest using 2045 as an endpoint year for calculating GWP | In order to meet 1.5° C peak warming target in line with Paris Accord |

3. **Assumption #2: Directional and Temporal Fungibility of Carbon across Sources and Sinks**

A key basis for carbon offsetting is the concept that one unit of $CO_2$ has the same climatic impact emitted anywhere on Earth, since $CO_2$ is a well-mixed GHG. However, this does not apply in the reverse direction for the same amount of $CO_2$ absorbed by different ecosystem types and in different forms. There are numerous natural sources and sinks of carbon that vary in their residence times and by location; therefore, they vary in the magnitude of their effects on climate.[45] The time that the average carbon molecule is stored in a natural carbon sink or "reservoir" before being emitted back into the atmosphere as $CO_2$ is the mean residence time, which can be calculated by dividing the reservoir carbon content in kg by the net carbon flux out of the reservoir in kg/year.[46] Within a given ecosystem, soil organic carbon, woody biomass, and non-woody biomass (e.g., leaf litter) typically vary in their individual carbon residence times.

Forestry projects such as the protection or expansion of forested areas are one of the most common approaches to offset $CO_2$ emissions,[47] operating under the assumption that their carbon storage can be reliably predicted. However, controlling the time that carbon remains in a natural carbon sink like woody biomass is complicated by chemical, physical, and biological forces. Reported estimates for the mean residence time of carbon in forest woody biomass range from 12 to 200 years,[48,49] and this value can be variable or uncertain even for the same ecosystem types (Figure 2).[49] Additionally, determining how increased atmospheric concentrations of $CO_2$ and higher global temperatures affect carbon residence time in forests is an active area of research,[50] and even related conditions like water availability can affect the carbon residence time of trees.[51] This complex uncertainty suggests a low likelihood that carbon credits from different forestry projects will have the same storage lifetime and, therefore, the same ultimate contribution to climate goals.

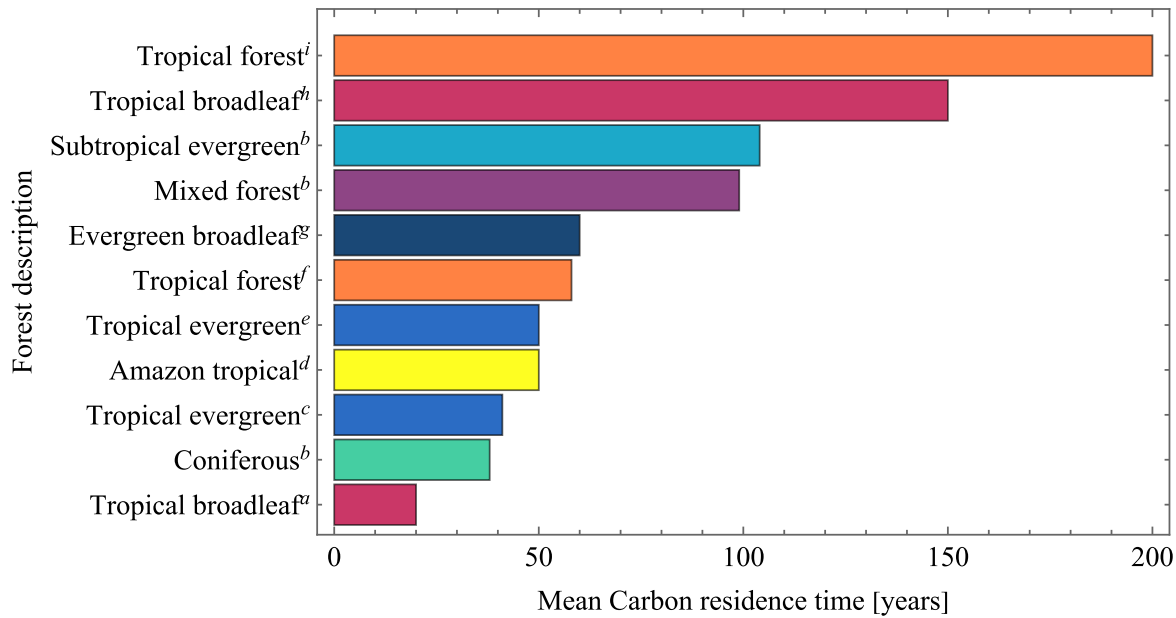

**Figure 2**: Mean residence times for carbon in woody biomass in different forest ecosystems described by: [a]Arain et al (2006), [b]Zhang et al (2010), [c]Kohlmaier (1997), [d]Hirsch et al (2004), [e]Post et al (1997), [f]Schaefer et al (2008), [g]Wang et al (2010), [h]Warnant et al (1994), [i]Kaduk et al (1996). [49,52–61]

Furthermore, the range in potential carbon residence time in woody biomass does not match the lifetime of $CO_2$ released into the atmosphere and thus does not approximate a permanent sink of $CO_2$. Long-term abatement of emissions through forestry carbon offsets cannot be quantified with certainty, even with attentive management of these projects. A forest carbon project assessed in India was found to have sequestered just 37% of the carbon estimated up to the first verification period, while another forest project in India only sequestered 3%.[62] Forest canopy cover in Northern India was found not to have increased after decades of tree-planting initiatives in the region.[63] Disasters and unplanned disturbances can also affect the longevity of carbon offset projects. For example, in July 2021, forests in Oregon in which Green Diamond Timber was paid to slow logging activities in exchange for carbon credits were burned down by the Bootleg fire.[64] The reforestation initiative at Mt Elgon National Park, Uganda, reforested only 8,000 out of 25,000 planned hectares before the project shut down 31 years early due to civil conflicts.[65]

Even systems of international support and management for carbon offsets have not resulted in improved outcomes of such projects. A group of countries participate in a voluntary climate change framework called Reducing Emissions from Deforestation in Developing countries (REDD) (now REDD+, with the plus sign standing for additional forest-related activities), which was established in 2007. An analysis of REDD+ projects in Cambodia revealed instances of carbon credits being generated for projects that never materialized.[66] Even Norway, the largest funder of REDD+, has acknowledged that projected results had been long delayed, and the risk of logging remained high.[67] By 2018, payments to five countries had been delayed at least five years as projects could not be verified. In 2021, Indonesia terminated its reforestation deal with Norway over failure to receive compensation.[68]

These are not just isolated incidents, but examples of irresolvable issues with trading carbon as a commodity. There have been innumerable failures in carbon trading, from the collapse of entire emissions trading systems to individual offset projects neglecting to deliver promised results for a variety of reasons.[69,70] Among these, errors in accounting and projects whose benefits cannot be verified are plentiful. In Canada, Alberta's emissions trading systems was declared a failure in 2011 as none of the agricultural credits could be verified.[71] Early in the development of the European Union Emissions Trading System, 170 million credits were mistakenly allocated, which went unnoticed and rewarded major polluters before disrupting the market with surplus credits.[72] The way that geographic borders are set up by the California Air Resources Board lets developers take advantage of mixed forest types being lumped together in transition areas.[73] Systemic over-crediting was found in California's forest offset program, inadvertently producing incentives to generate credits that do not represent genuine emissions reductions.[74] A forest offset project in New Mexico earned millions in carbon credits primarily due to being located where an erroneously low national average had been set.[73] The reforestation initiative at Mt Elgon National Park, Uganda, was originally projected to sequester at least 5,500 kg $CO_2$ hectare$^{-1}$ year$^{-1}$, a rate that seemingly omits the effects of plant respiration on net carbon uptake.[73]

4. **Assumption #3: Decreases in GHG Emissions are Fungible with GHGs Sequestered**

The accounting involved in carbon trading aims to fund improvements from business-as-usual scenarios, whether they lead to negative, halted, or only slower rates of GHG emissions. Carbon offsets include both decreases in projected GHG emissions and sequestration of GHGs, which ultimately produce different effects towards climate change mitigation. Decreasing emissions relative to expected future emissions does not reduce atmospheric GHG concentrations, but instead simply slows their continued growth. Unless the new GHG emission rate is zero, this approach produces a net increase in atmospheric GHGs over time, whereas actual sequestration does not. However, both prevent some amount of GHGs from being in the atmosphere into the future, which represents the quantity of carbon offset.

Ensuring that carbon offsets are at least associated with a verifiable reduction in GHG emissions is critical to carbon markets generating net benefits. The quality of carbon offsets depends on a long carbon residence time indicating 'permanence' and additionality, although there is no consensus in the exact criteria by which to evaluate carbon offsets. Carbon offsets are considered 'additional' only if the project would not have occurred without funding from purchased credits. Still, a large number of established projects have not been deemed additional. In a study on a sample of 12 projects in the Brazilian Amazon, only one contributed any additional reductions in deforestation, and 40% of the 50,000 tradable offset credits issued associated with that project were not genuinely additional.[75] Another report concluded that less than 10% of carbon capture projects meet criteria for high quality offsets likely to provide additional emissions reductions.[76] Similarly, a 2023 Guardian analysis concluded that over 90% of rainforest-based carbon credits verified by the world's leading certifier did not represent genuine reductions because of issues including lack of verifiable additionality and overinflated baselines.[77]

Quantifying reductions in emissions requires the establishment of a baseline rate of GHG emissions. Higher reductions are achievable when a baseline is artificially increased, leading to an exploitable loophole in carbon offsetting. For example, the projects designed to abate HFC-23 and $SF_6$ in Russia actually increased their waste gas generation levels to historically unprecedented amounts in order to generate credits.[78] Two types of waste gas projects, incinerating HFC-23 from HFC-22 and destroying $N_2O$ from adipic acid production, were found to account for 0.3% of registered projects but generated roughly half of the 1.5 billion credits issued up to that point because of the extremely high GWP associated with these GHGs.[78] Such practices that involve setting a baseline for a non-$CO_2$ GHG species like a refrigerant result in net increases of $CO_2$ *and* the refrigerant when a project is not truly additional and/or the baseline not accurate. In these cases, the oversimplification of the warming equivalency of different GHGs contributes to the risk of unintentionally worsening climate change via emissions trading.[79]

The support for carbon offsetting is not emergent from climate sciences, but instead largely from the mainstream economics literature, which seemingly misrepresents the efficacy of trading $CO_2$eq as a scientific consensus.[80,81] In reality, there have been calls to abandon carbon markets as a failed experiment,[82] not only due to difficulties in setting accurate baselines and verifying additionality for carbon offsets in practice,[83] but also due to ethical concerns that support the case against commodifying carbon.[84] Implemented carbon offset markets have been exploitative and undermine local control of resources.[85] The methods in place for equating GHG values can lead to unjust outcomes and incentivize delaying real solutions to climate change by making it cheaper for polluters in rich countries to pay developing nations not to utilize their natural resources.[13] This exacerbates existing inequalities while failing to provide meaningful achievements.

## 5. Policy Recommendations

### 5.1 Increase educational and collaborative opportunities for decision-makers at all levels

Communicating the intricate details of climate science has been an ongoing challenge that has no simple solution. Still, it is imperative that policymakers be made aware of several key concepts that contest the three simplifying assumptions that are ubiquitous in climate policy discussed above. Academic institutions can contribute to developing more effective climate policy through collaborations at all levels. Many such collaborations are possible without additional funding, and others would require new or diverted funding sources. Undergraduate research projects or theses can contribute valuable information for GHG inventories, for example, through case study analyses that only require data be made available. Graduate-level researchers in related fields can partner with organizations and agencies through more internship and fellowship opportunities. Science advisor positions already exist at the State and Federal levels in the US; this can be expanded more throughout local governments and agencies. Conferences and workshops can also help better disseminate up-to-date information. Better scientific literacy not only leads to more comprehensive climate policy, but also improves policymakers' ability to communicate with their constituents on crucial issues.[86] As a complimentary measure, climate scientists can be better informed on policy to help foster these collaborations.[87] Because these proposed policy changes involve deeper incorporation of the nuances and complexities of climate change, they are more likely to lead to effective climate change mitigation. They may also lead to greater stakeholder engagement through partnerships. Furthermore, decision-makers must incorporate this knowledge into their proposed pathways in order to avoid worsening the crisis through misguided abatement efforts and to plan genuinely effective climate change mitigation.

### 5.2 Separate emissions targets, i.e., the 'multi-basket' approach

We also recommend that policies set individual emissions reduction targets by GHG species to more appropriately reflect how different gases contribute to climate change. This 'multi-basket' approach is most conducive to achieving temperature stabilization targets, particularly with limited or no overshoot.[32,35,88] Setting a target of net-zero $CO_2$ without relying on the abatement benefits of SLCFs encourages a focus on major technological and systemic change, as opposed to the incremental changes that have so far failed to bend the emissions curve. Instead of a simple net-zero goal and aggregated approach to GHG accounting, policy-makers will need to make more nuanced decisions and detailed GHG targets, as is already part of the GHG Protocol for the agriculture industry.[89] With the 'multi-basket' approach, previously neglected pollutants like black carbon and carbon monoxide can also be more appropriately monitored and reduction goals set.[26]

The 'multi-basket' approach introduces new opportunities as different sectors recognize their roles in meeting specific targets that are more tangible to them than an overall lump-sum net-zero goal for which the responsibility is spread across all industries. For example, aiming to reduce the rate of methane emissions to stabilize its warming effect might be more realistic for planning for methane-intensive sectors including natural gas, waste management, and ruminant agriculture. Comparing the reported aggregated to disaggregated emissions for food production shows that nearly half of the carbon footprint of beef and the majority of the carbon footprint of rice is actually methane in terms of $CO_2eq$ on the 100-year time frame.[90] The choice of emission metric and temporal horizon also influences the comparison between different types of beef production; the carbon footprint of grass-fed beef is higher than non-grass-fed beef using the 100-year GWP, but lower when using the 100-year GTP to aggregate GHG emissions to $CO_2eq$.[91]

This breaking down of a decarbonization goal into its GHG components also produces a first step in taking a target from a goal to an actionable plan. Communicating such GHG-specific mitigation goals to constituents may further inform the general public of the various types and sources of GHGs and improve general climate literacy.

### 5.3 Reduced or avoided emissions that depend on a baseline should not generate offset credits

Avoiding carbon offsets that depend on a baseline or 'business as usual' scenario is necessary to prevent economic exploitation of climate change mitigation. This practice has already been observed, where baseline scenarios are grossly inflated to generate offset credits on the premise of then 'reducing' these supposed activities that generate emissions.[78] Carbon credits generated by reducing or avoiding emissions never actually remove GHGs from the atmosphere, and whenever the assumed baseline is higher than the real historical level of emissions, there is actually an increase from 'business as usual'.

Nature-based solutions may have many co-benefits to human and ecosystem health, but cannot reliably generate tradeable carbon credits, because the residence times of carbon in ecosystems is never as permanent as the fossil carbon being 'offset.' These types of projects also tend to depend on a baseline for comparison that involve speculation, for example, preserving forests that would have otherwise been logged.[92] There are multiple approaches to setting a baseline to calculate carbon credits and Griscom et al 2009 found that the amount of credited emissions can range over two orders of magnitude depending on which baseline approach is used.[93]

### 5.4 Prioritize immediate and permanent carbon capture projects with the lowest life cycle impacts, but not as equal offsets

Net-zero decarbonization targets that include offsetting emissions should carefully differentiate between temporarily and permanently sequestered carbon and prioritize offsetting projects that are permanent, such as direct air capture with geological sequestration. However, there are dozens of startups selling carbon credits to build and expand their operations that represent carbon they will eventually remove from the atmosphere once fully scaled up, and there is a high likelihood that not all startups will reach their intended scale at their planned pace. There are simply too many companies in the space to succeed long term, many using technologies that have yet to be demonstrated at scale. Selling credits preemptively is necessary to develop and grow these start-ups, which allows polluters to believe and act as if emissions are being offset as they are purchased, instead of theoretically in the future. This is another irresolvable issue with trading carbon credits on a voluntary market. According to CDR.fyi, an organization tracking offset purchases and carbon dioxide removal, less than 3% of the more than 28 million tons of $CO_2$ sold has actually been delivered as the time of this writing.[94] As of May 2025, Climeworks, one of the most established CDR companies founded in 2009, had still not successfully scaled up enough to capture more $CO_2$ than its operations emit, and the price of $CO_2$ per ton is 10 times the target price it had for this year.[95] This means that from an absolute emissions standpoint, the company has yet to demonstrate a net climate benefit.

There are also many different methods for capturing and storing carbon from the atmosphere, that utilize different materials and technologies, most of which have not been investigated to understand their carbon footprints and net climate benefits, which depends on several factors including what type of fuel is used and the ultimate use of the captured $CO_2$.[96] An accurate understanding of the geospatial carbon footprint of different methods of CDR will demonstrate which are viable for climate change mitigation and where, and which technologies and locations to avoid. Without thorough studying, it is unclear which carbon dioxide removal methods being deployed are resulting in more carbon emitted than removed, especially when the end-use is considered for captured carbon being incorporated into commercial products.

## 6. Conclusion

In this study, we clarified a set of simplifications in climate science that must be understood by decision-makers. First, different GHGs are not perfectly fungible using GWP, and there are notable uncertainties involved in comparing the climate change contributions of different GHGs. Next, time and direction matter in quantifying the GHG emissions reduction or sequestration potential of various options. The choice of temporal horizon for normalizing non-$CO_2$ GHGs to $CO_2$eq mass units strongly influences the relative impact of different gases, especially for SLCFs including methane and industrial refrigerants. The residence time of carbon sequestered in different biological sinks ranges from days to many decades depending on the location, type of sink, and management. Carbon residence times can vary by orders of magnitude across sinks and forms, from days to millennia. Last, reductions in rates of GHG emissions are not equivalent to carbon sequestration, just as slowing down is not the same as reversing course. Based on the failures of carbon offsetting inherent to the carbon trading market, we recommend more sophisticated net-zero policies that do not heavily rely on carbon credits to succeed. More oversight is possible for actions directly undertaken by a governing body instead. Avoiding carbon offsets may generate an additional benefit of decoupling climate change mitigation from monetary value. We aimed to provide an accessible overview of these concepts in this paper for policymakers as a first step, but more communication and collaboration between researchers and decision-makers at all levels is needed.

**Acknowledgments:**
The authors thank two funding sources from the National Science Foundation (NSF): Alyssa Pfadt-Trilling's NSF Graduate Research Fellowship and NSF Award #2316124. The author would like to recommend the following resources for readers who would like to learn more on the issues discussed in this paper: Barbara Haya and Emily Clayton's repository of articles on carbon offset quality: https://gspp.berkeley.edu/research-and-impact/centers/cepp/projects/berkeley-carbon-trading-project/repository-of-articles and the ongoing research at the Berkeley Carbon Trading Project (https://gspp.berkeley.edu/research-and-impact/centers/cepp/projects/berkeley-carbon-trading-project)

**Author Contributions:**

*A.P-T.:* Conceptualization, methodology, formal analysis, investigation, data curation, visualization, writing - original draft. *M-O.F.:* Project administration, supervision, writing - review and editing.

**Competing Interests:**

The authors have no competing interests to declare.